\documentclass[12pt]{article}
\def\1{\'{\i}}
\def\3{\ss}
\def\e{\, {\rm e}}

\usepackage[cm]{fullpage}

\title{Approximate Spacetime for Neutron Stars}
\author{Francisco Frutos-Alfaro \\ 
\small{School of Physics, Space Research Center} \\ \small{and} \\ 
\small{Laboratory of Theoretical Physics and Computation of 
the University of Costa Rica}}
\date{\today}

\begin{document}
\maketitle

\abstract{
An approximate realistic metric representing the spacetime of neutron stars is 
obtained by perturbing the Kerr metric. This metric has five parameters, namely 
the mass, spin or angular momentum, mass quadrupole, spin octupole and 
mass hexadecapole. Moreover, a version of the Hartle-Thorne metric containing 
these parameters is constructed by means of a series transformation between 
these spacetimes and solving the Einstein field equations. The form of the 
Pappas metric in Schwarzschild spherical coordinates is found. The three 
relativistic multipole structures are compared.}

\section{Introduction}

\noindent
Among compact objects are neutron stars (NS). These stars are relativistic 
rotating objects with high density, and strong gravitational and magnetic 
fields. The study of NS is relevant to understand the extrem conditions of 
matter in there, the behaviour of particles around them, and the structure 
of its spacetime \cite{LP,Camenzind,Berti1}.  

\noindent
The quest to find a realistic spacetime representation for neutron stars (NS) 
is an important task in astrophysics. Many attemps to obtain this spacetime 
have been done from approximate metrics until numerical metrics. The first 
attemp was made by Hartle-Thorne (HT). The relevance of the HT work was that 
they matched the interior solution with the exterior one \cite{HT}. 
Quevedo and Mashhoon \cite{QM} and Manko and Novikov \cite{MN} obtained exact 
solutions with charge and arbitrary mass multipoles. Later, more exact 
solutions containing other features, for instance magnetic dipole, were found 
\cite{Pachon}. With the advent of computer technology, the implementations of 
computer programs to find numerical solutions become a vogue \cite{Stergioulas}.
However, approximate solutions are still important to extract astrophysical 
information from NS \cite{Pappas}. Moreover, a fourth order HT metric 
for the exterior of neutron stars was obtained in \cite{Yagi}.

\noindent
There are several techniques to find solutions of the Einstein field equations 
(EFE). Among them, the Ernst formalism has played an important role in 
finding new exact and approximate solutions. This formalism is employed in 
\cite{QM,MN,Pappas}. In this contribution, however, we include features like 
mass quadrupole, spin octupole and mass hexadecapole to the Kerr metric 
perturbatively. This is achieved easily by means of perturbing the Lewis metric 
form of the Kerr spacetime \cite{Frutos1,Frutos2}. The form of the 
perturbations due to spin octupole and mass hexadecapole has the structure 
proposed by Ryan \cite{Ryan}. Then, one is certain that these features are 
introduced in the right manner. This metric has the advantage that it reduces 
to the Kerr metric which is an exact solution with mass and angular momentum. 
Moreover, it is simple to implement computationaly. 

\noindent
The HT spacetime does not have spin octupole, mass hexadecapole, and the 
interactions mass-quadrupole, quadrupole-quadrupole, and spin-quadrupole. 
Nevertheless, this metric gives excelent results of the inner most stable 
circular orbit (ISCO) of particles around NS \cite{Berti2}. Adding these 
interactions, and the spin octupole and the mass hexadecapole to this metric 
would improve considerably its applicability in computational calculations. 
One can guess an approximate HT metric with these features from our deformed 
Kerr spacetime, by finding a transformation between them from the post-linear 
forms of these metrics and solving the EFE. Both spacetime were tested to be 
solutions of the EFE by means of REDUCE programs, and these programs 
are available upon request. 

\noindent
This paper is organized as follows. The perturbation method of the Kerr metric 
using the Lewis spacetime is discussed in section 2. In Section 3, the 
construction of the HT version is presented via a series transformation for the 
first order in spin octupole and mass hexadecapole. The inclusion of the 
interactions of the spin octupole and mass hexadecapole with the mass, spin, 
quadrupole and with each other is also found in this section. The relativistic 
multipole structure is found using the Fodor-Hoenselaers-Perj\'es method 
\cite{Fodor} in Section 4. In Section 5, the Pappas metric \cite{Pappas} is 
compared with the ones presented here. 
In the last section some conclusions are presented.

\section{Generating the Metric}

\noindent
The Lewis metric \cite{Lewis} was successfully applied to find approximate 
solutions of the EFE using the Erez-Rosen metric and the Kerr 
as seed metrics \cite{Frutos1,Frutos2}. It is given by

\begin{eqnarray}
\label{Lewis}
d {s}^{2} = - V d {t}^{2} + 2 W d {t} d {\phi} 
+ X d {\rho}^{2} + Y d {z}^{2} + Z d {\phi}^{2} 
\end{eqnarray}

\noindent
where the chosen canonical coordinates are $ x^{1} = \rho $ and 
$ x^{2} = z $. The potentials $ V, \, W, \,  X, \, Y $ and $ Z $ are functions 
of $ \rho $ and $ z $ with $ \rho^{2} = V Z + W^{2} $.

\noindent
The transformation that leads to the Kerr metric is

\begin{eqnarray}
\label{mapp}
\rho = \sqrt{\Delta} \sin{\theta} \quad {\rm and} \quad 
z = (r - m) \cos{\theta} , 
\end{eqnarray}

\noindent
where $ \Delta = r^2 - 2 m r + a^2 $, $ m $ and $ a $ are the mass and 
the rotational parameter. The angular momentum or spin is given by 
$ {\cal S}_1 = J = m a $. 

\noindent
The Ansatz for the Lewis potentials to include the spin octupole parameter, 
$ S_{3} $ and the mass hexadecapole parameter, $ M_{4} $, is

\begin{eqnarray}
\label{potentials}
V & = & V_{K} \, {\rm e}^{- 2 \psi} , \nonumber \\
W & = & W_K + W_{ms} , \nonumber \\
X & = & (X_K + X_{ms}) \, {\rm e}^{2 \chi} , \\
Y & = & (Y_K + Y_{ms}) \, {\rm e}^{2 \chi} , \nonumber \\
Z & = & (Z_K + Z_{ms}) \, {\rm e}^{2 \psi} , \nonumber 
\end{eqnarray} 

\noindent
where the potentials $ V_K, \, W_K, \, X_K, \, Y_K, \, Z_K $ are the Lewis 
potentials for the Kerr metric. The perturbation terms are 
$ W_{ms}, \, X_{ms}, \, Y_{ms}, \, Z_{ms} $ which include terms with $ M_4 $ and 
$ S_3 $, and interaction of these parameters with the other ones. 

\noindent
The functions $ \psi, \, \chi, \, W_{ms}, \, X_{ms}, \, Y_{ms} $, and $ Z_{ms} $ 
are chosen as follows

\begin{eqnarray}
\label{Anstaz}
\psi & = & \psi_q + \psi_{ms} , \nonumber \\ 
\chi & = & \chi_q + \chi_{ms} , \nonumber \\
W_{ms} & = & \left(\xi_0 \frac{J q}{r^{4}} 
+ \xi_1 \frac{S_{3}}{r^{3}} \right) h_{31}
+ \xi_2 \frac{m S_3}{r^4} h_{32} + \xi_3 \frac{q S_3}{r^6} h_{33} 
+ \xi_4 \frac{M_4 S_3}{r^8} h_{34} + \xi_5 \frac{J M_4}{r^6} h_{35} 
, \nonumber \\
X_{ms} & = & \mu_1 \frac{m M_4}{r^6} h_{41} + \mu_2 \frac{q M_4}{r^8} h_{42} 
+ \mu_3 \frac{M_4^2}{r^{10}} h_{43} , \\
Y_{ms} & = & \iota_1 \frac{m M_4}{r^4} h_{51} + \iota_2 \frac{q M_4}{r^6} h_{52}
+ \iota_3 \frac{M_4^2}{r^8} h_{53} + \iota_4 \frac{S_3^2}{r^6} h_{54} , 
+ \iota_5 \frac{J S_3}{r^4} h_{55} , \nonumber \\
Z_{ms} & = & \left(\zeta_1 \frac{m M_4}{r^4} h_{51} 
+ \zeta_2 \frac{q M_4}{r^6} h_{52} + \zeta_3 \frac{M_4^2}{r^8} h_{53} 
+ \zeta_4 \frac{S_3^2}{r^6} h_{54} + \zeta_5 \frac{J S_3}{r^4} h_{55} 
\right) \sin^2{\theta} , \nonumber
\end{eqnarray}

\noindent
where 

\begin{eqnarray}
\label{Anstaz2}
\psi_q & = & \frac{q}{r^3} P_2 + 3 \frac{m q}{r^4} P_2 , \nonumber \\ 
\chi_q & = & \frac{q P_2}{r^3} 
+ \frac{1}{3} \frac{m q}{r^4} (5 P_2^2 + 5 P_2 - 1) 
+ \frac{1}{9} \frac{q^2}{r^6} (25 P_2^3 - 21 P_2^2 - 6 P_2 + 2) ,  
\nonumber \\
\psi_{ms}  & = & \gamma_1 \frac{M_{4}}{r^{5}} h_{11}
+ \gamma_2 \frac{m M_4}{r^6} h_{12} + \gamma_3 \frac{q M_4}{r^8} h_{13} 
+ \gamma_4 \frac{S_3^2}{r^8} h_{14} + \gamma_5 \frac{J S_3}{r^6} h_{15} 
, \nonumber \\
\chi_{ms} & = & \eta_1 \frac{M_{4}}{r^{5}} h_{21}
+ \eta_2 \frac{m M_4}{r^6} h_{22} + \eta_3 \frac{q M_4}{r^8} h_{23} 
+ \eta_4 \frac{M_4^2}{r^{10}} h_{24} + \eta_5 \frac{S_3^2}{r^8} h_{25} 
+ \eta_6 \frac{J S_3}{r^6} h_{26} . \nonumber
\end{eqnarray}

\noindent
In \cite{Frutos3}, $ \psi_q $ and $ \chi_q $ were found. The functions 
$ h_{ij} $ are functions of $ \theta $ only. The first order terms of 
$ \psi_{ms} , \, \chi_{ms} $ and $ W_{ms} $ are taken from \cite{Ryan}. 
The term $ J q $ of $ W_{ms} $ represents the interaction of the spin with 
the quadrupole, which was not considered in \cite{Frutos3}. 

\noindent
From this Ansatz, the perturbative terms can be determined by solving the EFE.
The functions $ h_{ij} $ are combinations of Legendre polynomials, 
$ P_n(\cos{\theta}) ,\, n = 1, \dots, \, 8 $, and an associated Legendre 
polynomial, $ P^1_3(\cos{\theta}) = (5 P_2 + 1) \sin{\theta} . $  
After solving the EFE, the functions $ h_{ij} $ are given by

\begin{eqnarray}
\label{hs}
h_{11} & = & {P_{4}} , \nonumber \\
h_{12} & = & {P_{4}} , \nonumber \\
h_{13} & = & {P_{7}} , \nonumber \\
h_{14} & = & \frac{7}{64} + \frac{7}{48} P_2 + \frac{63}{352} P_4 
+ \frac{175}{528} P_6 + \frac{16}{429} P_7 , \nonumber \\
h_{15} & = & - \frac{3}{4} P_2 - P_4 , \nonumber \\
h_{21} & = & {P_{4}} , \nonumber \\
h_{22} & = & \frac{45}{11} {P_{4}} , \nonumber \\
h_{23} & = & {P_{7}} , \nonumber \\
h_{24} & = & - \left(\frac{5}{99} + \frac{625}{9009} P_2 + \frac{90}{1001} P_4
+ \frac{250}{1683} P_6 + \frac{12250}{24453} P_8 
\right) , \nonumber \\
h_{25} & = & - \frac{287}{1728} - \frac{91}{528} P_2 - \frac{7}{416} P_4 
+ \frac{175}{144} P_6 + \frac{16}{429} P_7 , \\
h_{26} & = & \frac{5}{4} P_2 - 3 P_4 , \nonumber \\
h_{31} & = & {P^{1}_{\,3}} \sin{\theta} , \nonumber \\
h_{32} & = & \frac{5}{2} P_2 - \frac{8}{63} P_3 - \frac{5}{2} P_4 
+ \frac{8}{63} P_5 , \nonumber \\
h_{33} & = & - \frac{1}{10} - \frac{1}{6} P_2 - \frac{27}{55} P_4 
+ \frac{25}{33} P_6 , \nonumber \\
h_{34} & = & - \frac{5}{33} P_2 - \frac{25}{143} P_4 - \frac{35}{66} P_6 
+ \frac{245}{286} P_8 , \nonumber \\
h_{35} & = & \frac{20}{11} (P_4 - P_6) , \nonumber \\
h_{41} & = & {P_{3}} , \nonumber \\
h_{42} & = & - \frac{24}{7} P_2 + \frac{600}{77} P_4 + \frac{300}{143} P_6 
, \nonumber \\
h_{43} & = & {P_{4}} , \nonumber \\
h_{51} & = & \frac{2}{33} - \frac{2}{5} P_1 + \frac{10}{33} P_2 
+ \frac{2}{5} P_3 + \frac{6}{11} P_4 , \nonumber \\
h_{52} & = & - \frac{3}{26} + \frac{47}{26} P_2 P_4 - \frac{303}{182} P_2 
- \frac{5}{182} P_4 , \nonumber \\
h_{53} & = & \frac{33191}{4157010} + \frac{33191}{831402} P_2 
+ \frac{165259}{692835} P_4 + \frac{882}{3553} P_6 + \frac{882}{2717} P_8 , 
\nonumber \\
h_{54} & = & \frac{125293}{247104} + \frac{35315}{82368} P_2 
- \frac{581}{9152} P_4 - \frac{6125}{6336} P_6 , \nonumber \\ 
h_{55} & = & 3 (P_4 - P_2) . \nonumber 
\end{eqnarray}

\noindent
The constants are found to be 

\begin{eqnarray}
\gamma_1 & = & 1 , \quad \gamma_2 = 5 , \quad 
\gamma_3 = \gamma_4 = \gamma_5 = 1 , \nonumber \\
\eta_1 & = & \eta_2 = \eta_3 = \eta_4 = \eta_5 = \eta_6 = 1 , \nonumber \\
\xi_0 & = & 1 , \quad \xi_1 = \frac{7}{12} , \quad 
\xi_2 = \xi_3 = \xi_4 = \xi_5 = 1 , \nonumber \\
\mu_1 & = & 4, \quad \mu_2 = \mu_3 = 1 , \nonumber \\
\iota_1 & = & \iota_2 = \iota_3 = \iota_4 = \iota_5 = 1 , \nonumber \\
\zeta_1 & = & \zeta_2 = \zeta_3 = \zeta_4 = \zeta_5 = - 1 . \nonumber
\end{eqnarray}
 
\noindent
The new metric potentials are

\begin{eqnarray}
\label{newpotentials}
V & = & \frac{1}{{\Sigma}^2} \left[\Delta - a^2 \sin^2{\theta} \right] \, 
{\rm e}^{- 2 \psi} , \nonumber \\
W & = & - \frac{2 J r}{{\Sigma}^2} \sin^2{\theta} + W_{ms} , \nonumber \\
X & = & \left(\frac{{\Sigma}^2}{\Delta} + X_{ms} \right) \, {\rm e}^{2 \chi} , \\
Y & = & \left({\Sigma}^2 + Y_{ms} \right) \, {\rm e}^{2 \chi} , \nonumber \\
Z & = & \left(\frac{1}{{\Sigma}^2} 
\left[(r^2 + a^2)^2 - a^2 \Delta \sin^2{\theta} \right] 
- Y_{ms} \right) \, {\rm e}^{2 \psi} \sin^2{\theta} , \nonumber 
\end{eqnarray} 

\noindent
where $ {\Sigma}^2 = r^2 + a^2 \cos^2{\theta} $.

\noindent
This metric is valid up to third order in all parameters, 
including the interactions of all parameters with each other.

\noindent
A post-linear expansion of the metric can be written as

\begin{eqnarray}
\label{potentials2}
V & \simeq & \left(1 - 2 U - 2 \frac{m a^2}{r^3} \cos^2{\theta} 
\right) \, {{\rm e}^{- 2 \psi}} , \nonumber \\
W & \simeq & - 2 \frac{J}{r} \sin^2{\theta} + W_{ms} , \\
X & \simeq & \left(1 + 2 U + 4 U^2 - \frac{a^2}{r^2} \sin^2{\theta} 
- 2 \frac{m a^2}{r^3} (1 + \sin^2{\theta}) 
- 4 \frac{m^2 a^2}{r^4} (2 + \sin^2{\theta}) + X_{ms} \right) 
{{\rm e}^{2 \chi}} , \nonumber \\
Y & = & r^2 \left(1 + \frac{a^2}{r^2} \cos^2{\theta} + \frac{Y_{ms}}{r^2} \right) 
\, {{\rm e}^{2 \chi}} , \nonumber \\
Z & \simeq & r^2 \sin^2{\theta} \left(1 + \frac{a^2}{r^2} 
+ 2 \frac{m a^2}{r^3} \sin^2{\theta} - \frac{Y_{ms}}{r^2} \right) 
\, {{\rm e}^{2 \psi}} , \nonumber 
\end{eqnarray} 

\noindent
where $ U = {m}/{r} $. In \cite{Frutos3} is the complete expansion without 
the $ S_3 $ and $ M_4 $ terms.

\section{Constructing a new Hartle-Thorne Metric}

\noindent
Adding perturbatively some features, for example the spin octupole and 
the mass hexadecapole to the HT metric would be interesting, because this 
metric is still used as a comparison with more realistic metrics. 
The HT metric is an approximate solution of the EFE with three parameters, 
mass, angular momentum and mass quadrupole. It is given by

\begin{eqnarray}
\label{HT}
d {s}^2 & = & - F_1 d t^2 + F_2 d R^2 
+ R^2 F_3 \left[d \theta^2 + \sin^2{\theta} (d \phi - \omega d t)^2 \right] \\
& = & - V_{HT} d t^2 - 2 W_{HT} dt d \phi + X_{HT} d R^2 
+ R^2 Y_{HT} \left[d \theta^2 + \sin^2{\theta} d \phi^2 \right] , \nonumber  
\end{eqnarray}

\noindent
where

\begin{eqnarray}
\label{funcs}
F_1 & = & \left(1 - 2 U + 2 \frac{J^2}{R^4} \right) 
\left[1 + 2 K_1 P_2 \right] \nonumber \\ 
& \simeq & \left(1 - 2 U + 2 \frac{J^2}{R^4} \right) \e^{2 \psi_1} , \nonumber \\
F_2 & = & \left[1 - 2 K_2 P_2 \right] 
\left({1 - 2 U + 2 \frac{J^2}{R^4}} \right)^{-1} \\ 
& \simeq & \e^{- 2 \psi_2} 
\left({1 - 2 U + 2 \frac{J^2}{R^4}} \right)^{-1} , \nonumber \\ 
F_3 & = & 1 - 2 K_3 P_2 \simeq \e^{- 2 \psi_3} , \nonumber \\
\omega & = & 2 \frac{J}{R^3} . \nonumber 
\end{eqnarray}

\noindent
The functions $ K_1, K_2 $, and $ K_3 $ are given by

\begin{eqnarray}
\label{ks}
K_1 & = & \frac{J^2}{m R^3} (1 + U) 
+ \frac{5}{8} \left(\frac{q}{m^3} - \frac{J^2}{m^4} \right) 
Q^{2}_{2} \left(\frac{R}{m} - 1 \right) , \nonumber \\
K_2 & = & K_1 - 6 \frac{J^2}{R^4} = K_1 - 6 \frac{J^2}{m^4} U^4 , \\
K_3 & = & \left(K_1 + \frac{J^2}{R^4} \right) 
+ \frac{5}{4} \left(\frac{q}{m^3} - \frac{J^2}{m^4} \right) 
\frac{U}{\sqrt{1 - 2 U}} Q^{1}_{2} \left(\frac{R}{m} - 1 \right) , \nonumber
\end{eqnarray}

\noindent
where $ U = M/R $. The functions $ Q^{1}_{2} $ and $ Q^{2}_{2} $ are 
Legendre functions of the second kind

\begin{eqnarray}
\label{lfsk}
Q^{1}_{2} (x) = \sqrt{x^2 - 1} \left(\frac{3}{2} x 
\ln{\left(\frac{x + 1}{x - 1} \right)} 
- \frac{(3 x^2 - 2)}{(x^2 - 1)} \right) , \nonumber \\
Q^{2}_{2} (x) = ({x^2 - 1}) \left(\frac{3}{2} 
\ln{\left(\frac{x + 1}{x - 1} \right)} 
- \frac{(3 x^3 - 5 x)}{(x^2 - 1)^2} \right) . \nonumber
\end{eqnarray}

\noindent
The metric potencials are

\begin{eqnarray}
\label{potht}
V_{HT} & \simeq & \left(1 - 2 U - \frac{2}{3} \frac{J^2}{R^4} \right) 
\, {{\rm e}^{2 \alpha_1}} , \nonumber \\
W_{HT} & = & - 2 \frac{J}{R} \sin^2{\theta} , \nonumber \\
X & \simeq & \left({1 - 2 U + 2 \frac{J^2}{R^4}} \right)^{-1} 
\, {{\rm e}^{- 2 \alpha_2}} , \\
Y_{HT} & \simeq & {{\rm e}^{- 2 \alpha_3}} , \nonumber 
\end{eqnarray} 

\noindent
where 

\begin{eqnarray}
\label{psis}
\alpha_1 & = & \left(K_1 + \frac{4}{3} \frac{J^2}{R^4} \right) P_2 , \nonumber \\
\alpha_2 & = & K_2 P_2 , \\
\alpha_3 & = & K_3 P_2 . \nonumber
\end{eqnarray} 

\noindent
The Taylor expansion of $ K_1, \, K_2 $ and $ K_3 $ are

\begin{eqnarray}
\label{ks2}
K_1 & = & \frac{q}{R^3} + 3 \frac{m q}{R^4} - 2 \frac{J^2}{R^4} , \nonumber \\
K_2 & = & \frac{q}{R^3} + 3 \frac{m q}{R^4} - 8 \frac{J^2}{R^4} , \nonumber \\
K_3 & = & \frac{q}{R^3} + \frac{5}{2} \frac{m q}{R^4} 
- \frac{1}{2} \frac{J^2}{R^4} . \nonumber
\end{eqnarray}

\noindent
The complete expansion of the HT including the second order terms in $ q $ was 
found in \cite{FS,Frutos3}.

\noindent
To guess an improvement of the HT metric, we have to find a solution of the EFE 
compatible with HT metric. In order to do it, we will propose an Ansatz. 
In \cite{FS}, the second order in $ q $ for the post-linear HT was found 
perturbatively. A transformation that converts the post-linear Kerr-like metric 
(\ref{potentials2}) without $ S_3 $ and $ M_4 $ into an improved HT was 
obtained in \cite{Frutos3}. The same transformation can be used to transform 
the post-linear Kerr-like metric (\ref{potentials2}) with $ S_3 $ and $ M_4 $ 
at first order into an improved HT in the post-linear form of (\ref{potht}) 
with $ S_3 $ and $ M_4 $ at first order, changing $ q \rightarrow m a^2 - q $.
This transformation is \cite{Frutos3}

\begin{eqnarray}
\label{trans}
r & = & R \left[1 + \frac{m q}{R^4} f_1 + \frac{q^2}{R^6} f_2 
+ \frac{a^2}{R^2} \left({h_1} + \frac{m}{R} h_2 + \frac{m^2}{R^2} h_3 \right) 
\right] , \\
\theta & = & \Theta + \frac{m q}{R^4} g_1 + \frac{q^2}{R^6} g_2 
+ \frac{a^2}{R^2} \left({h_4} + \frac{m}{R} h_5 \right) , \nonumber
\end{eqnarray}

\noindent
where

\begin{eqnarray}
\label{functs}
f_1 & = & \frac{1}{9} (5 P_2^2 - 4 P_2 - 1) , \nonumber \\
f_2 & = & \frac{1}{72} (40 P_2^3 - 24 P_2^2 - 43) , \nonumber \\
g_1 & = & \frac{1}{6} (2 - 5 P_2) \cos{\Theta} \sin{\Theta} , \nonumber \\
g_2 & = & \frac{1}{6} P_2 (2 - 5 P_2) \cos{\Theta} \sin{\Theta} , \\
h_1 & = & - \frac{1}{2} \sin^2{\Theta} , \nonumber \\
h_2 & = & - \frac{1}{2} \sin^2{\Theta} , \nonumber \\
h_3 & = & 1 - 3 \cos^2{\Theta} = - 2 P_2 , \nonumber \\
h_4 & = & - \frac{1}{2} \cos{\Theta} \sin{\Theta} , \nonumber \\
h_5 & = & - \cos{\Theta} \sin{\Theta} , \nonumber 
\end{eqnarray}

\noindent
with $ P_2 = P_2(\cos{\Theta}) $

\noindent
Now, considering this fact, the $ J q $ interaction term and the terms due to 
the spin octupole $ S_3 $ and the mass hexadecapole $ M_4 $ at second order, 
the Ansatz of an improved HT metric functions is from (\ref{potentials2})

\begin{eqnarray}
\label{pothtnew}
V & \simeq & \left(1 - 2 U - \frac{2}{3} \frac{J^2}{R^4} \right) 
\, {{\rm e}^{2 \psi_1}} , \nonumber \\
W & \simeq & - \left[2 \frac{J}{R} + \left(\frac{J q}{R^{4}} 
- \frac{7}{12}\frac{S_{3}}{R^{3}} \right) (5 P_2 + 1) \right] \sin^2{\Theta} 
+ {\tilde W}_{ms}, 
\nonumber \\
X & \simeq & \left({1 - 2 U + 2 \frac{J^2}{R^4}} \right)^{-1} 
\, {{\rm e}^{- 2 \psi_2}} , \\
Y & \simeq & R^2 \, {{\rm e}^{- 2 \psi_3}} , \nonumber \\
Z & \simeq & R^2 \, {{\rm e}^{- 2 \psi_3}} \sin^2{\Theta} , \nonumber 
\end{eqnarray} 


\noindent
where

\begin{eqnarray}
\label{psinew}
\psi_1 & = & \
\frac{q}{R^3} P_2 + 3 \frac{m q}{R^4} P_2 
- \frac{2}{3} \frac{J^2}{R^4} P_2 + \frac{M_4}{R^5} P_4 + \psi_{1 ms} , \\
\psi_2 & = & \frac{q}{R^3} P_2 + 3 \frac{m q}{R^4} P_2
+ \frac{1}{24} \frac{q^2}{R^6} [16 P^2_2 + 16 P_2 - 77] 
- 8 \frac{J^2}{R^4} P_2 + \frac{M_4}{R^5} P_4 + \psi_{2 ms} , \nonumber \\
\psi_3 & = & \frac{q}{R^3} P_2 + \frac{5}{2} \frac{m q}{R^4} P_2 
+ \frac{1}{72} \frac{q^2}{R^6} [28 P^2_2 - 8 P_2 + 43]
- \frac{1}{2} \frac{J^2}{R^4} P_2 + \frac{M_4}{R^5} P_4 + \psi_{3 ms} , \nonumber
\end{eqnarray}

\noindent
with

\begin{eqnarray}
\label{funcsnew}
\psi_{1 ms} & = & \frac{m M_4}{r^6} {\tilde h}_{12} 
+ \frac{q M_4}{r^8} {\tilde h}_{13} + \frac{J S_3}{r^6} {\tilde h}_{14} 
+ \frac{S_3^2}{r^8} {\tilde h}_{15} , \nonumber \\
\psi_{2 ms} & = & \frac{m M_4}{r^6} {\tilde h}_{22} 
+ \frac{q M_4}{r^8} {\tilde h}_{23} + \frac{J S_3}{r^6} {\tilde h}_{24} 
+ \frac{M_4^2}{r^{10}} {\tilde h}_{25} + \frac{S_3^2}{r^8} {\tilde h}_{26} , \\
\psi_{3 ms} & = & \frac{m M_4}{r^6} {\tilde h}_{32} 
+ \frac{q M_4}{r^8} {\tilde h}_{33} + \frac{J S_3}{r^6} {\tilde h}_{34} 
+ \frac{M_4^2}{r^{10}} {\tilde h}_{35} + \frac{S_3^2}{r^8} {\tilde h}_{36} ,
\nonumber \\
{\tilde W}_{ms} & = & \frac{m S_3}{r^4} {\tilde h}_{42}
+ \frac{q S_3}{r^6} {\tilde h}_{43} + \frac{M_4 S_3}{r^8} {\tilde h}_{44}
+ \frac{J S_3}{r^5} {\tilde h}_{45} + \frac{J M_4}{r^6} {\tilde h}_{46} . 
\nonumber
\end{eqnarray}

\noindent
The $ {\tilde h}_{ij} $ are found solving the EFE perturbatively. These 
functions are 

\begin{eqnarray}
\label{hs2}
{\tilde h}_{12} & = & 3 P_4 , \nonumber \\
{\tilde h}_{13} & = & P_7 , \nonumber \\
{\tilde h}_{14} & = & \frac{3}{4} P_2 + P_4 , \nonumber \\
{\tilde h}_{15} & = &  - \frac{7}{48} P_2 - \frac{63}{352} P_4 
- \frac{175}{528} P_6 - \frac{7}{64} , \nonumber \\
{\tilde h}_{22} & = & 5 P_4 - \frac{110}{27} , \nonumber \\
{\tilde h}_{23} & = & - \frac{20}{21} P_2 + \frac{36}{77} P_4 
+ \frac{16}{33} P_6 + P_7 - 7 , \nonumber \\
{\tilde h}_{24} & = & - \frac{15}{4} P_2 + \frac{11}{2} P_4 - 5 , \nonumber \\
{\tilde h}_{25} & = & \frac{1000}{693} P_2 + \frac{360}{1001} P_4 
+ \frac{20}{99} P_6 + \frac{280}{1287} P_8 + \frac{5}{18} , \nonumber \\
{\tilde h}_{26} & = & - \frac{679}{144} P_2 - \frac{1505}{1056} P_4 
- \frac{1085}{792} P_6 - \frac{637}{96} , \nonumber \\
{\tilde h}_{32} & = & \frac{245}{27} P_2^2 - \frac{70}{27} P_2 - 1 , \nonumber \\
{\tilde h}_{33} & = & \frac{5}{21} P_2 + \frac{12}{77} P_4 
+ \frac{47}{132} P_6 + P_7 + 1 , \\
{\tilde h}_{34} & = & \frac{7}{3} P_4 + 1 , \nonumber \\
{\tilde h}_{35} & = & - \frac{50}{693} P_2 + \frac{45}{1001} P_4 
+ \frac{8}{99} P_6 + \frac{217}{1287} P_8 , \nonumber \\
{\tilde h}_{36} & = & \frac{259}{576} P_2 - \frac{35}{264} P_4 
- \frac{10255}{12672} P_6 + 1 , \nonumber \\
{\tilde h}_{42} & = & \frac{5}{2} (P_2 - P_4) , \nonumber \\
{\tilde h}_{43} & = & \frac{1}{6} P_2 + \frac{27}{55} P_4 
- \frac{25}{33} P_6 + \frac{1}{10} , \nonumber \\
{\tilde h}_{44} & = & \frac{5}{33} P_2 + \frac{25}{143} P_4 
+ \frac{35}{66} P_6 - \frac{245}{286} P_8 , \nonumber \\
{\tilde h}_{45} & = & 0 , \nonumber \\
{\tilde h}_{46} & = & 4 P_2 P_4 - \frac{8}{7} P_2 - \frac{20}{7} P_4 . \nonumber 
\end{eqnarray}

\noindent
This metric is solution of the EFE up to the third order in all parameters 
($ m, \, q, \, M_4, \, J, \, S_3 $).

\section{Relativistic Multipole Moments}

To determine if two metric are isometric, one has to compare its multipole 
structure. It is useful to find this structure for our spacetime. 
At first glance, our metric has 5 complex multipoles ($ {\cal M}_0 = m, \, 
{\cal S}_1 = J = m a, \, {\cal M}_2 = q - m a^2, \, {\cal S}_3, 
\, {\rm and} \, {\cal M}_4) $. To see if it is true, one has to 
construct the Ernst potential for this metric. This potential is given by 
\cite{Ernst}

\begin{eqnarray}
\label{Ernst}
{\cal E} = f + i \Omega ,
\end{eqnarray}

\noindent
where $ f = V = V_K {\rm e}^{- 2 \psi} $ and $ \Omega $ is the twist scalar. 
To get this scalar, the following equation has to be solved

\begin{eqnarray}
\label{twisteq}
\partial_{\alpha} \Omega = 
\varepsilon_{\alpha \beta \mu \nu} k^{\beta} \nabla^{\mu} k^{\nu} ,
\end{eqnarray}

\noindent
where $ k^{\beta} $ is the Killing vector, $ \nabla^{\mu} $ is the contravariant 
derivative and 
$ \varepsilon_{\alpha \beta \mu \nu} = \sqrt{- g} \epsilon_{\alpha \beta \mu \nu} $
($ g $ is determinant of the metric tensor). Let us take the Killing vector as 
in the Kerr metric $ k^{\beta} = (1, \, 0, \, 0, \, 0) $. 
Then, the approximate solution of (\ref{twisteq}) is

\begin{eqnarray}
\label{twist}
\Omega = - 2 \frac{J}{\rho^2} \cos{\theta} + {\cal H} ,
\end{eqnarray}

\noindent
where

\begin{eqnarray}
\label{funcom}
{\cal H} = \frac{S_3}{r^4} h_{61} 
+ \frac{J q}{r^5} h_{62} + \frac{m S_3}{r^5} h_{63} 
+ \frac{q S_3}{r^7} h_{64} 
+ \frac{J M_4}{r^7} h_{65} 
+ \frac{M_4 S_3}{r^9} h_{66} , 
\end{eqnarray}

\noindent
with

\begin{eqnarray}
\label{hmom}
h_{61} & = & \frac{7}{12} (5 P_2 - 2) \cos{\theta} , \nonumber \\
h_{62} & = & 4 P_2 \cos{\theta} , \nonumber \\
h_{63} & = & \frac{1}{420} (490 (5 P_2 - 2) \cos{\theta} - 96 P_4) , \\
h_{64} & = & \frac{1}{6} (4 P_2 - 18 P_4 - 7) \cos{\theta} , \nonumber \\
h_{65} & = & 4 P_4 \cos{\theta} , \nonumber \\
h_{66} & = & \frac{1}{66} (- 110 P_2 + 54 P_4 - 175 P_6) \cos{\theta} . 
\nonumber
\end{eqnarray}

\noindent
Now, the Ernst function is given by

\begin{eqnarray}
\label{functer}
\xi = \frac{1 + {\cal E}}{1 - {\cal E}} .
\end{eqnarray}

\noindent
It is easy to show that this Ernst function and its inverse are solutions of 
the Ernst equation \cite{Ernst}

$$ (\xi \xi^{\star} - 1) \nabla^2 \xi = 2 \xi^{\star} [\nabla \xi]^2 . $$

\noindent
To calculate the relativistic multipole moments, it is better to employ the 
inverse function \cite{Ernst}. Moreover, it is custumary to employ the prolate 
spheroidal coordinates $ (t, \, x, \, y, \, \phi) $. 
The transformation to these coordinates is achieved by means of

\begin{eqnarray}
\label{prolate}
\sigma x & = & r - M , \\
y & = & \cos{\theta} , \nonumber
\end{eqnarray}

\noindent
where $ \sigma^2 = M^2 - a^2 $.

\noindent
The method to obtain the relativistic multipole moments is the following 
\cite{Fodor}

\begin{enumerate}
\item use the inverse Ernst function $ \xi^{-1} $ in prolate coordinates,
\item set $ y = \cos{\theta} = 1 $ into $ \xi^{-1} $,
\item change $ \sigma x \rightarrow 1/z $ into $ \xi^{-1} $,
\item expand in Taylor series of $ z $ the inverse Ernst function, and finally,
\item employ the Fodor-Hoenselaers-Perj\'es (FHP) formulae \cite{Fodor}.
\end{enumerate}

\noindent
A REDUCE program that calculates the multipole moment was written with 
this recipe. The first ten complex moments 
$ {\cal P}_{n} = {\cal M}_n + i {\cal S}_n $ are

\begin{eqnarray}
\label{sk}
{\cal P}_0 & = & {\cal M}_0 = m , \nonumber \\
{\cal P}_1 & = & i {\cal S}_1 = i J = i m a , \nonumber \\
{\cal P}_2 & = & {\cal M}_2 = q - m a^2 , \nonumber \\
{\cal P}_3 & = & i {\cal S}_3 
= - i \left(m a^3 + \frac{7}{8} S_3 \right) , \nonumber \\
{\cal P}_4 & = & {\cal M}_4 + i {\cal S}_4 = m a^4 + M_4 
+ i \left(- 2 J q + \frac{4}{35} m S_3 \right) , \nonumber \\
{\cal P}_5 & = & {\cal M}_5 + i {\cal S}_5 
= \frac{7}{4} J S_3 - 4 m M_4 + i m a^5 , \\
{\cal P}_6 & = & {\cal M}_6 + i {\cal S}_6 
= - m a^6 + i \left(\frac{7}{4} q S_3 - 2 J M_4 \right) , \nonumber \\
{\cal P}_7 & = & {\cal M}_7 + i {\cal S}_7 =
\left(\frac{1}{5} q M_4 + \frac{16}{429} S_3^2 \right) 
- i m a^7 , \nonumber \\
{\cal P}_8 & = & {\cal M}_8 + i {\cal S}_8 = 
m a^8 + i \frac{7}{4} M_4 S_3 , \nonumber \\
{\cal P}_9 & = & i {\cal S}_9 = i m a^9 , \nonumber \\
{\cal P}_{10} & = & {\cal M}_{10} = - m a^{10} . \nonumber 
\end{eqnarray}

\noindent
The real parts are the massive multipoles, $ {\cal M}_{i} $ and 
the imaginary parts are the spin multipoles, $ {\cal S}_{i} $. 
If one eliminates mixed terms and $ S_3^2 \sim 0 $ in (\ref{sk}) the multipole 
structure becomes simpler. 

\noindent
For neutron stars, the form of the first five multipole moments 
are \cite{Pappas,Yagi,Pappas2}

\begin{eqnarray}
\label{neutron}
{\cal M}_0 & = & m , \nonumber \\
{\cal S}_1 & = & S = J = m a , \nonumber \\
{\cal M}_2 & = & - \alpha {m a^2} , \\
{\cal S}_3 & = & - \beta {m a^3} , \nonumber \\
{\cal M}_4 & = & \gamma {m a^4} , \nonumber 
\end{eqnarray}

\noindent
where $ \alpha, \, \beta, \, {\rm and} \, \gamma $ are parameters.

\noindent
It is easy to see that if one sets

\begin{eqnarray}
\label{val}
q & = & (1 - \alpha) m a^2 , \nonumber \\
S_3 & = & - \frac{7}{8} (\beta + 1) m a^3 , \\ 
M_4 & = & (\gamma - 1) m a^4 , \nonumber
\end{eqnarray}

\noindent
a similar multipole structure is obtained from (\ref{sk}).  

\noindent
Now, let us determine the multipole structure of the new HT spacetime. 
The twist scalar for this HT metric is

\begin{eqnarray}
\label{twistht}
\Omega & = & \left[- 2 J u^2 + \frac{7}{12} S_3 (5 P_2 - 2) u^4  
+ \frac{1}{6} (- 24 J q P_2 + 7 M S_3 (5 P_2 - 2)) u^5 \right. \\
& + & \left. \frac{1}{6} (- 24 J M_4 P_4 + q S_3 (- 4 P_2 + 18 P_4 + 7)) u^7 
+ \frac{1}{66} M_4 S_3 (110 P_2 - 54 P_4 + 175 P_6) u^9  
\right] \cos{\theta} . \nonumber 
\end{eqnarray}

\noindent
After using (\ref{pothtnew}) and (\ref{twistht}) to construct the Ernst 
functions, we find that the relativistic multipole moments for this HT metric 
are

\begin{eqnarray}
\label{rmsht}
{\cal M}_0 & = & M , \nonumber \\
{\cal S}_1 & = & J , \nonumber \\
{\cal M}_2 & = & - q , \\
{\cal S}_3 & = & - \frac{7}{8} S_3 , \nonumber \\
{\cal M}_4 & = & M_4 . \nonumber 
\end{eqnarray}

\noindent
Obviously, from this multipole structure, it is possible to calculate the 
multipole moments of a neutron star, as well.

\section{Comparison with the Pappas Metric}

\noindent
Pappas found an approximate solution of the EFE by means of the Ernst method 
\cite{Pappas}. This spacetime has 5 parameters 
$ M, \, M_2, \, M_4, \, S_1 = J, \, S_3 $, which, by construction, represent 
the relativistic multipole moments. The metric is given in cylindrical 
Weyl-Papapetrou coordinates by

\begin{eqnarray}
\label{pappassp}
d s^2 = - f (d t - \omega d \phi)^2 
+ \frac{1}{f} \left[ {\rm e}^{2 \gamma} (d \rho^2 + d z^2) 
+ \rho^2 d \phi^2 \right] , 
\end{eqnarray}

\noindent
where

\begin{eqnarray}
\label{papmet}
f & = & 1 - 2 \frac{M}{\eta} + 2 \frac{M^2}{\eta^2} 
+ \frac{1}{\eta^5} \left[(M_2 - M^3) \rho^2 - 2 (M^3 + M_2) z^2 \right] 
\nonumber \\
& + & \frac{1}{\eta^6} \left[2 z^2 (M^4 - J^2 + 2 M M_2) 
- 2 M M_2 \rho^2 \right] + \frac{A}{28 \eta^9} 
+ \frac{B}{14 \eta^{10}} , \nonumber \\
\omega & = & - 2 \frac{J}{\eta^3} \rho^2 - 2 \frac{M J}{\eta^4} \rho^2
+ \frac{F}{\eta^7} + \frac{H}{2 \eta^8} + \frac{G}{4 \eta^{11}} , \\
\gamma & = & \frac{1}{4 \eta^8} \rho^2 \left[J^2 (\rho^2 - 8 z^2) 
+ M (M^3 + 3 M_2) (\rho^2 - 4 z^2) \right]
- \frac{M^2}{2 \eta^4} \rho^2 , \nonumber
\end{eqnarray}

\noindent
with

\begin{eqnarray}
\label{papvals}
\eta & = & \sqrt{\rho^2 + z^2} , \nonumber \\
A & = & 8 \rho^2 z^2 (24 M J^2 + 17 M^2 M_2 + 21 M_4) 
+ \rho^4 (7 M^5 - 10 M J^2 + 32 M^2 M_2 - 21 M_4) \nonumber \\
& + & 8 z^4 (- 7 M^5 + 20 M J^2 - 22 M_2 M^2 - 7 M_4) , \nonumber \\
B & = & \rho^4 (10 M^2 J^2 + 10 M^3 M_2 + 21 M M_4 + 7 M_2^2) \nonumber \\
& + & 4 z^4 (7 M^6 - 40 M^2 J^2 - 14 J S_3 + 30 M^3 M_2 + 14 M M_4 + 7 M_2^2) 
\nonumber \\
& - & 4 \rho^2 z^2 (7 M^6 + 27 J^2 M^2 - 21 J S_3 + 48 M^3 M_2 + 42 M M_4 
+ 7 M_2^2) , \nonumber \\
F & = & \rho^4 (S_3 - M^2 J) - 4 \rho^2 z^2 (M^2 J + S_3) , \\
G & = & \rho^2 (- J^3 (\rho^4 + 8 z^4 - 12 \rho^2 z^2) 
+ M J ((M^3 + 2 M_2) \rho^4 \nonumber \\
& - & 8 (3 M^3 + 2 M_2) z^4 + 4 (M^3 + 10 M_2) \rho^2 z^2) 
+ M^2 S_3 (3 \rho^4 - 40 z^4 + 12 \rho^2 z^2)) , \nonumber \\
H & = & 4 \rho^2 z^2 (J (M_2 - 2 M^3) - 3 M S_3) + \rho^4 (J M_2 + 3 M S_3) .
\nonumber
\end{eqnarray}

\noindent
To see which form has this metric in spherical-like coordinates, 
we use the Kerr mapping (\ref{mapp}) with $ a = 0 $. 
Then, the function $ \eta^2 $ is

\begin{eqnarray}
\label{eta}
\eta^2 & = & \Delta + (M^2 - a^2) \cos^2{\theta} 
= r (r - 2 M) + M^2 \cos^2{\theta} . 
\end{eqnarray}

\noindent
Substituting (\ref{eta}) in the metric functions (\ref{papmet}) and 
expanding in Taylor series up to $ {\cal O}(r^{-6}) $, 
the metric potentials take the form 

\begin{eqnarray}
\label{pappot}
V & = & f \nonumber \\  
& = & 1 - 2 M u - 2 M_2 P_2 u^3 
- \frac{2}{3} \left[J^2 (2 P_2 + 1) + 3 M M_2 P_2 \right] u^4 \nonumber \\
& + & \frac{1}{63} \left[- 3 M J^2 (14 P_2^2 + 20 P_2 + 14) 
- M M_2 (70 P_2^2 + 88 P_2 - 14) \right. \nonumber \\
& + & \left. 7 M_4 (- 35 P_2^2 + 10 P_2 + 7) \right] u^5 , \nonumber \\  
W & = & f \omega \nonumber \\  
& = & \left[- 2 J u - \frac{2}{3} S_3 u^3 (5 P_2 + 1) 
+ \frac{1}{3} \left[3 J M_2 - 5 M S_3 \right] u^4 (5 P_2 + 1) \right.
\nonumber \\  
& + & \left. \frac{1}{3} \left[J^3 (- 7 P_2^2 + 8 P_2 + 5) 
+ 6 M J M_2 (- 3 P_2^2 + 7 P_2 + 2) \right. \right. \nonumber \\  
& + & \left. \left. M^2 S_3 (7 P_2^2 - 50 P_2 - 11) \right] u^5 
\phantom{\frac{1}{3}} \!\!\!\!\! \right] \sin^2{\theta} , \nonumber \\ 
X & = & \frac{1}{f \Delta} 
[(r - M)^2 \sin^2{\theta} + \Delta \cos^2{\theta}] \, {\rm e}^{2 \gamma} 
\nonumber \\  
& = & 1 + 2 M u + 4 M^2 u^2 + 2 (4 M^3 + M_2 P_2) u^3 \nonumber \\ 
& + & \frac{2}{3} \left[24 M^4 + 3 J^2 P_2^2 
+ M M_2 (5 P_2^2 + 11 P_2 - 1) \right] u^4 \\  
& + & \frac{1}{63} \left[2016 M^5 + 3 M J^2 (266 P_2^2 - 36 P_2 - 14) 
+ M^2 M_2 (1330 P_2^2 + 1096 P_2 - 266) \right. \nonumber \\  
& + & \left. 7 M_4 (35 P_2^2 - 10 P_2 - 7) \right] u^5 , \nonumber \\  
Y & = & \frac{1}{f} 
[(r - M)^2 \sin^2{\theta} + \Delta \cos^2{\theta}] \, {\rm e}^{2 \gamma} 
\nonumber \\  
& = & r^2 \left[1 + 2 M_2 u^3 P_2  
+ \frac{2}{3} \left[3 J^2 P_2^2 + M M_2 (5 P_2^2 + 5 P_2 - 1) \right] u^4 
\right. \nonumber \\  
& + & \left. \frac{1}{63} \left[3 M J^2 (182 P_2^2 - 36 P_2 - 14) 
+ M^2 M_2 (910 P_2^2 + 172 P_2 - 182) \right. \right. \nonumber \\    
& + & \left. \left. 7 M_4 (35 P_2^2 - 10 P_2 - 7) \right] u^5 
\phantom{\frac{2}{3}} \!\!\!\!\! \right] , \nonumber \\  
Z & = & \frac{\rho^2}{f} - f \omega^2 \nonumber \\  
& = & r^2 \sin^2{\theta} \left[1 + 2 M_2 P_2 u^3 
+ 2 (J^2 (2 P_2 - 1) + 3 M M_2 P_2) u^4 \phantom{\frac{2}{3}} \right. 
\nonumber \\  
& + & \left. \frac{1}{63} \left[3 M J^2 (14 P_2^2 + 188 P_2 - 70) 
+ M^2 M_2 (70 P_2^2 + 844 P_2 - 14) \right. \right. \nonumber \\  
& - & \left.\left. 7 M_4 (35 P_2^2 - 10 P_2 - 7) \right] u^5 
\phantom{\frac{2}{3}} \!\!\!\!\! \right] , \nonumber 
\end{eqnarray}

\noindent
where $ u = 1 / r $. By means of a REDUCE program, we checked that the metric 
potentials fulfill the EFE. According to Pappas the multipole structure of his 
spacetime is $ {\cal M}_{2 n} = M, \, M_2, M_4 $ 
for $ n = 0, \, 1, \, 2 $ and $ {\cal S}_{2 n + 1} = S_1, \, S_3 $ for 
$ n = 0, \, 1 $. The twist potential for the expanded Pappas spacetime is

\begin{eqnarray}
\label{fompap}
\Omega = - \frac{1}{3} \left[ 6 J u^2 + S_3 u^4 (10 P_2 - 4) 
+ u^5 \left(12 J M_2 P_2 + M S_3 (20 P_2 - 8) \right) \right] 
\cos{\theta} . 
\end{eqnarray}

\noindent
Using (\ref{pappot}) and (\ref{fompap}) in our program, 
the multipole structure is as expected

\begin{eqnarray}
\label{papmult}
{\cal M}_0 & = & M , \nonumber \\
{\cal S}_1 & = & J , \nonumber \\
{\cal M}_2 & = & M_2 , \\
{\cal S}_3 & = & S_3 , \nonumber \\
{\cal M}_4 & = & M_4 . \nonumber 
\end{eqnarray}

\noindent
Then, our metric (\ref{potentials}) have not the same multipole structure, 
therefore they are not isometric. The multipole structures of the new HT and 
the Pappas metrics are similar. Setting $ q = - M_2 $ and rescaling $ S_3 $ 
in (\ref{rmsht}) both metrics become isometric. From (\ref{pappot}), 
the Pappas metric contains the post-linear versions of the Schwarzschild, 
Erez-Rosen (up to $ M M_2 $) and the Lense-Thirring metrics. 
Our metric contains the Kerr metric and the post-linear version of 
the Erez-Rosen metric (up to second order in $ q $).

\section{Conclusions}

\noindent
We have found an approximate solution of the EFE by means of perturbing the 
Kerr metric. This approximate solution has five parameters, mass, angular 
momentum, mass quadrupole, spin octupole and mass hexadecapole. 
The mass quadrupole, spin octupole and mass hexadecapole were included 
perturbatively. It is valid up to the third order in these parameters. 

\noindent
By finding the twist scalar, we found the multipole structure employing 
the FHP formalism. It is possible to choose the multipole parameters, 
so that the first five multipole moments of a neutron star are similar. 
The simple form of our spacetime does it easy to implement computationaly. 
Including more relativistic multipole moments to our metric is easy using 
the procedure described here. Our metric presents an advantage over the Pappas 
metric, because it does not contain the Kerr spacetime as a limiting case. 

\noindent
Through a transformation we guessed an improved HT metric at first order 
in the spin octupole and mass hexadecapole. Solving the EFE, a new version of 
the HT, including mixed and cuadratic terms was also found. This is important, 
because the original HT can be matched with interior solutions and is used to 
validate spacetimes. The twist scalars and the multipole structures were also 
found for this improved HT and the Pappas metric. A comparison reveals that 
this HT metric is isometric with the Pappas metric after doing a transformation.

\noindent
Our spacetime has potentialy many applications. It could be used to infer the 
properties of the structure of a neutron star from astrophysical observations.
Another task is to find the ISCO as a function of the mass, mass quadrupole, 
mass hexadecapole, spin and spin octupole, as an extention of \cite{Chaverri}. 
An interesting future work is to find an interior solution for our spacetime.

\end{document}